\documentclass[graybox]{svmult}

\usepackage{mathptmx}       
\usepackage{helvet}         
\usepackage{courier}        
\usepackage{type1cm}        
\usepackage{makeidx}         
\usepackage{graphicx}        
\usepackage{multicol}        
\usepackage[bottom]{footmisc}

\makeindex             

\begin{document}

\title*{Ground-Based {\boldmath $BVRI$} Follow-Up Observations of the Cepheid V1154 Cyg in {\it Kepler's} Field}
\author{C.-C. Ngeow, R. Szab{\'o}, L. Szabados, A. Henden, M.A.T. Groenewegen \& the {\it Kepler} Cepheid Working Group}
\authorrunning{Ngeow et al.} 
\institute{C.-C. Ngeow \at National Central University, Jhongli City, 32001, Taiwan. \email{cngeow@astro.ncu.edu.tw}
}
%
%
\maketitle

\vspace{-3.0cm}

\abstract*{The almost un-interrupted observations from {\it Kepler Space Telescope} can be used to search for Earth-size and larger planets around other stars, as well as for stellar variability and asteroseismological study. However, the {\it Kepler's} observations are carried out with a single broad-band filter, and ground-based follow-up observations are needed to complement {\it Kepler's} light curves to fully characterize the properties of the target stars. Here we present ground-based optical ($BVRI$) follow-up observations of V1154 Cyg, the only Cepheid in the {\it Kepler} field of view, and deriving basic properties of this star. }

\abstract{The almost un-interrupted observations from {\it Kepler Space Telescope} can be used to search for Earth-size and larger planets around other stars, as well as for stellar variability and asteroseismological study. However, the {\it Kepler's} observations are carried out with a single broad-band filter, and ground-based follow-up observations are needed to complement {\it Kepler's} light curves to fully characterize the properties of the target stars. Here we present ground-based optical ($BVRI$) follow-up observations of V1154 Cyg, the only Cepheid in the {\it Kepler} field of view, and deriving basic properties of this star. 
}

\vspace{-0.5cm}
\section{Introduction}\label{sec:1}

V1154 Cyg ($P=4.925454$ days) is a known Cepheid located within {\it Kepler's} field-of-view. Analysis of this Cepheid using {\it Kepler's} light curves has been published in \cite{sza11} (hereafter S11). Some of the ground based follow-up observations (including optical and spectroscopic observations) can be found in \cite{mol07,mol09} and S11. The aim of this work is to provide further details of the $BVRI$ follow-up for V1154 Cyg, to supplement S11. Details of observations and data reduction are given in S11.

\section{Results: Basic Properties of V1154 Cyg}\label{sec:1}

\begin{figure}
\sidecaption[t]
\includegraphics[scale=.25]{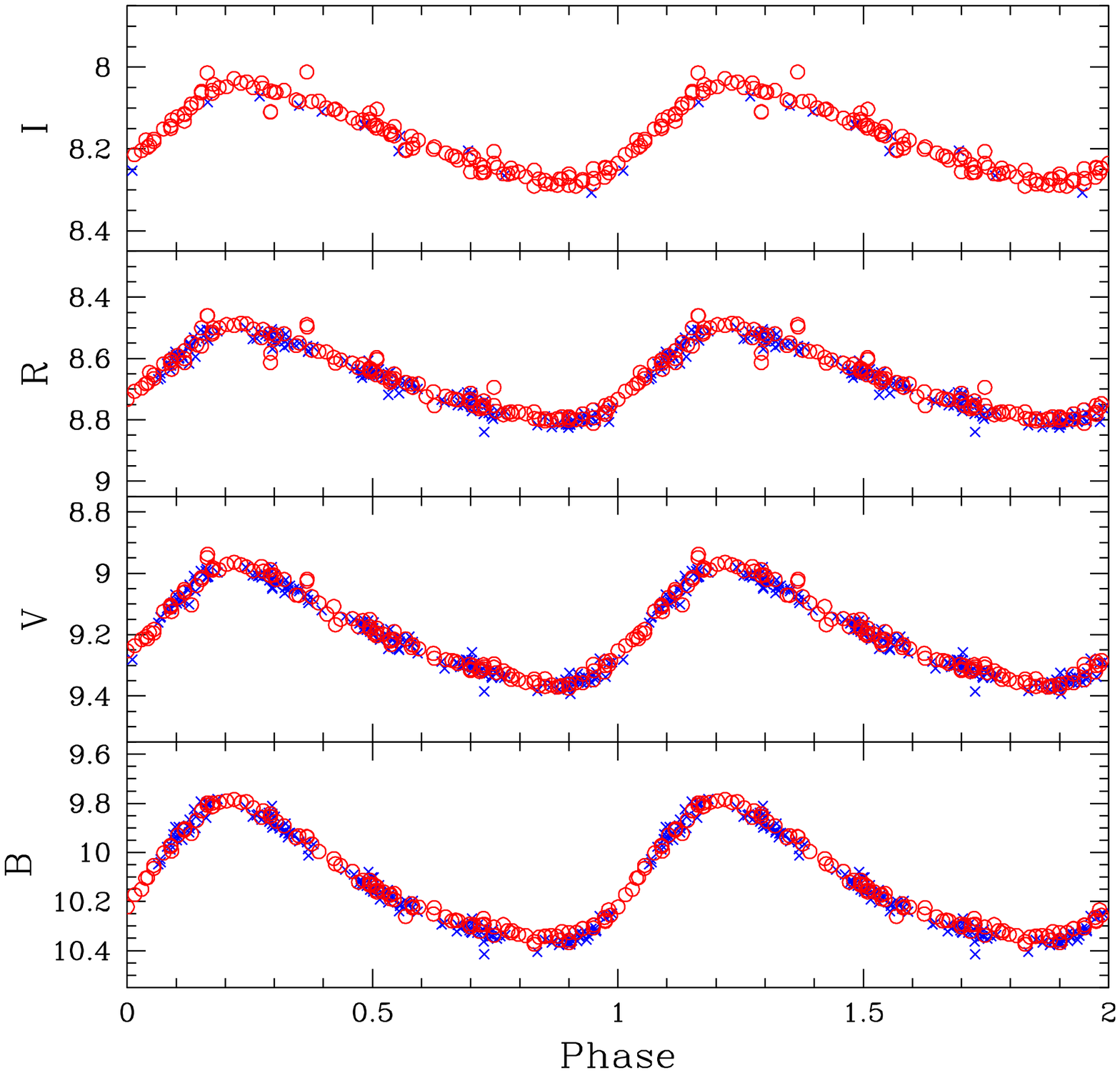}
\caption{Comparison of the $BVRI$ light curves from S11 (open circles) and Berdnikov (2008, crosses). Note that a correction of $+0.054$ mag needs to be added to S11 $B$ band data to bring into agreement between the two light curves.}
\label{fig_compare}
\end{figure}

\begin{figure}
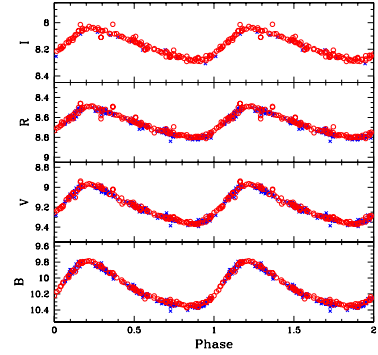
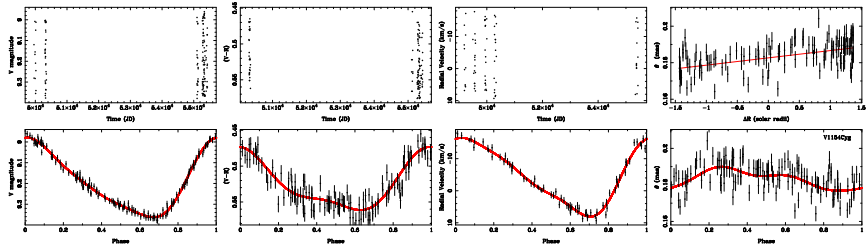

\includegraphics[scale=.2]{ngeowP_Fig2a.eps} 
\includegraphics[scale=.2]{ngeowP_Fig2b.eps} 
\includegraphics[scale=.2]{ngeowP_Fig2c.eps} 
\includegraphics[scale=.2]{ngeowP_Fig2d.eps} 
\caption{Results of the BW analysis for V1154 Cyg, including the fitting of $V$ band light curve, $(V-R)$ color curve, RV curve and angular diameters (from left to right).}
\label{fig_bw}
\end{figure}

$BVRI$ light curve properties and radial velocity measurements from spectroscopic observation suggested V1154 Cyg is a {\it bona fide} fundamental mode Cepheid. Figure \ref{fig_compare} compares our light curves to the light curves presented in \cite{ber08}. Table \ref{tab} summarizes the $BVRI$ intensity mean magnitudes and amplitudes (from Fourier fit to the light curves) based on S11 light curves. Baade-Wesselink (BW) surface brightness method was used to derive the distance and mean radius of V1154 Cyg, by combining the published radial velocity (RV) data and available light curves (details of the method can be found in \cite{gro08}). Figure \ref{fig_bw} presents the results of BW analysis. The distance and radius of V1154 Cyg are: $D = 1202\pm72\pm68$pc, $R/R_{\mathrm{sun}} = 23.5\pm1.4\pm1.3$, the first error is the formal fitting error, the second error is based on a Monte Carlo simulation taking into account the error in the photometry, RV data, $E(B-V)$ and the p-factor ($p = 1.255$ is adopted with a 5\% error).

\begin{table}
\caption{Basic $BVRI$ properties of V1154 Cyg.}
\label{tab}       
\begin{tabular}{p{4.5cm}p{1.5cm}p{1.5cm}p{1.5cm}p{1.5cm}}
\hline\noalign{\smallskip}
Band: & $B$ & $V$ & $R$ & $I$ \\
\noalign{\smallskip}\svhline\noalign{\smallskip}
Intensity mean magnitude & 10.107 & 9.185 & 8.659 & 8.168 \\
Amplitude                & 0.547  & 0.390 & 0.314 & 0.250 \\
\noalign{\smallskip}\hline\noalign{\smallskip}
\end{tabular}
\end{table}

\begin{acknowledgement}
CCN thanks the funding from National Science Council (of Taiwan) under the contract NSC 98-2112-M-008-013-MY3. The research leading to these results has received funding from the European Community’s Seventh Framework Programme (FP7/2007-2013) under grant agreement no. 269194 (IRSES/ASK). This project has been supported by the `Lend\"ulet' program of the Hungarian Academy of Sciences and the Hungarian OTKA grants K83790 and MB08C 81013. R.Sz. was supported by the J\'anos Bolyai Research Scholarship of the Hungarian Academy of Sciences.
\end{acknowledgement}


\vspace{-0.9cm}
\begin{thebibliography}{5}

\bibitem{sza11} Szab{\'o}, R., Szabados, L., Ngeow, C.-C., et al.\ 2011, MNRAS, 413, 2709 
\bibitem{mol07} Molenda-\.Zakowicz, J., Frasca, A., Latham, D. W. \& Jerzykiewicz, M. 2007, Acta Astronomica, 57, 301
\bibitem{mol09} Molenda-\.Zakowicz, J., Jerzykiewicz, M. \& Frasca, A. 2009, Acta Astronomica, 59, 213
\bibitem{ber08} Berdnikov, L.~N.\ 2008, VizieR On-line Data Catalog: II/285
\bibitem{gro08} Groenewegen, M.~A.~T.\ 2008, A \& A, 488, 25 


\end{thebibliography}
\end{document}